\documentclass{article}
\usepackage{ijcai25}

\usepackage[hidelinks]{hyperref}
\usepackage[utf8]{inputenc}
\usepackage[small]{caption}
\usepackage{graphicx}
\usepackage{amsmath}
\usepackage{amsthm}
\usepackage{booktabs}
\usepackage{algorithm}
\usepackage{algorithmic}
\usepackage{enumitem}
\usepackage{multirow}
\usepackage{algorithm}
\usepackage[algo2e,ruled,vlined]{algorithm2e}
\usepackage{graphicx}
\usepackage{amsmath}
\usepackage{booktabs}
\usepackage{multirow}
\usepackage{chngpage}
\usepackage{bbm}
\usepackage{enumitem}
\usepackage{subcaption}
\usepackage{xcolor}
\usepackage{lipsum}
\usepackage{tcolorbox}
\usepackage{times}
\usepackage{soul}
\usepackage{url}
\usepackage{amssymb}
\usepackage{balance}

\renewcommand{\eqref}[1]{\ref{#1}}

\title{Classifying and Tracking International Aid Contribution Towards SDGs}

\author{
Sungwon Park\footnotemark[1]$^{1,2}$
\and
Dongjoon Lee\footnotemark[1]$^{1}$
\and
Kyeongjin Ahn$^{1,2}$
\and
Yubin Choi$^{1}$
\and  \\
Junho Lee$^{3}$ 
\and 
Meeyoung Cha$^{2,1}$ 
\And
Kyung Ryul Park$^{1}$ 
\affiliations
$^1$  Korea Advanced Institute of Science and Technology (KAIST)  \\  $^2$  Max Planck Institute for Security and Privacy (MPI-SP)  \\
$^3$ United Nations (UN)\\
}

\begin{document}

\maketitle
\begin{abstract}
International aid is a critical mechanism for promoting economic growth and well-being in developing nations, supporting progress toward the Sustainable Development Goals (SDGs). However, tracking aid contributions remains challenging due to labor-intensive data management, incomplete records, and the heterogeneous nature of aid data. Recognizing the urgency of this challenge, we partnered with government agencies to develop an AI model that complements manual classification and mitigates human bias in subjective interpretation. By integrating SDG-specific semantics and leveraging prior knowledge from language models, our approach enhances classification accuracy and accommodates the diversity of aid projects. When applied to a comprehensive dataset spanning multiple years, our model can reveal hidden trends in the temporal evolution of international development cooperation. Expert interviews further suggest how these insights can empower policymakers with data-driven decision-making tools, ultimately improving aid effectiveness and supporting progress toward SDGs.
\looseness=-1
\end{abstract}

\section{Introduction}
International aid, commonly known as Official Development Assistance (ODA), refers to financial support provided by governments to help developing countries combat poverty and promote economic growth. For decades, international aid has played a crucial role in socioeconomic development, particularly in advancing the United Nations Sustainable Development Goals (SDGs), by providing essential resources to vulnerable populations~\cite{alsayyad2020linkages,hynes2013evolution}. Ensuring the effectiveness of aid and progress toward the SDGs requires rigorous tracking of funding allocation and impact assessment. 

The Creditor Reporting System (CRS), established by the Organization for Economic Cooperation and Development (OECD) in 1967, is the most authoritative and comprehensive database that records decades of data on international aid projects. This database provides consistent and reliable statistics on development finance and individual aid projects ~\cite{unaids2004creditor,morgenstern2022foreign}. The CRS is publicly accessible, ensuring transparency for policymakers, researchers, and civil society organizations (CSOs). 
Each year, OECD member countries are required to report project-level information on their aid activities, which includes 93 data features such as \textit{project description}, \textit{donor code}, and \textit{recipient code}, \textit{purpose code} (sector), and \textit{commitment value} (budget). CRS receives over 250,000 new project entries annually, detailing activities in various recipient countries. Since 2018, the system has asked donors to specify the relevant SDG (\textit{SDG focus index}) for each project to assess its alignment with global goals~\cite{dcd2018a,ola2018results}. However, this manual classification has lacked consistency and systematic rigor, owing to its labor-intensive, costly, and inherently subjective nature. Moreover, the absence of SDG-related records prior to 2018, coupled with the fact that over half of reported projects since then remain uncategorized, presents significant challenges to assessing aid effectiveness. These limitations hinder a comprehensive understanding of how aid impacts the SDGs.

Major development agencies, including the United Nations (UN) and OECD, are exploring methods to automate the classification of aid projects and link them to the SDGs ~\cite{ericsson2019connecting,pincet2019linking,lee:IJCAI}. However, these efforts are complex due to the heterogeneous nature of the aid projects contributed by various countries and agencies. This diversity leads to significant syntactic and semantic variations in the project descriptions in CRS. Subjective judgments and sociopolitical contexts further influence how projects are categorized under specific aid goal, often resulting in misalignment between project descriptions and their assigned categories. These inconsistencies hinder classifiers from generalizing to unseen projects, particularly those related to underrepresented SDGs. We validated these rationales through interviews with aid experts.

As part of a joint effort between government experts and AI researchers to automate aid categorization, we identified key demands and developed a multi-label text classifier to improve the mapping of SDG impact.  By integrating their semantic definitions into the classifier's encoder, our goal was to support aid management and targeting while minimizing the influence of human bias in subjective interpretation of aid projects.
Our classifier learns data semantics through self-supervised learning and leverages insights from the large language model (LLM) to incorporate pre-trained knowledge. The training objective is to improve generalization, particularly for underrepresented SDGs, by classifying previously unseen project description sentences during training.\footnote{Our code is made freely available to support broader use at \\\url{https://github.com/deu30303/ODA_SDG}}

We validated the AI classification results with aid experts and confirmed that our approach demonstrates exceptional performance in identifying relevant SDG categories. It effectively captures the complexities of real-world projects by filling in the past and missing data on a comprehensive scale. This capability also allows the analysis of aid financing trends from a longitudinal perspective, providing valuable information to policymakers. We present our findings from expert interviews on AI classification and its practical usefulness. 
Building on this achievement, we emphasize the need for a global effort to develop objective, transparent, and universally accepted standards for aid project classification through collaboration between aid experts and AI researchers.\looseness=-1

\section{Related Work}

\subsection{AI in the SDG Monitoring}
Tracking the progress of aid projects is essential for organizations such as the UN and the OECD. However, organizing development aid projects to align with the appropriate goals presents several challenges. These include: 1) the complexity of multi-label classification, which necessitates a comprehensive understanding of each project across 93 features and reasoning to map them to multiple SDG goals based on the 169 official targets; 2) interdependence and trade-offs between 17 SDGs; and 3) the imbalanced distribution of goals and data classes~\cite{ola2018results}. Previous research applied machine learning to categorize purpose codes within development aid projects ~\cite{ericsson2019connecting,lee:IJCAI,pincet2019linking}, leading to reduced expert labor and improved precision~\cite{lee:IJCAI}. However, previous approaches have been limited to single-label classification or have failed to account for the complexities of multi-label classification and semantic information, such as country-specific descriptions and SDG definitions, which are essential for accurately representing development aid projects. \looseness=-1

\subsection{Multi-Label Classification}

In multi-label text classification, several models have been proposed to capture label dependencies, including classifier chains~\cite{read2009classifier}, graph neural networks (GNNs)~\cite{zhang2018link}, and label-specific attention networks~\cite{xiao2019label}. Among them, classifier chains are susceptible to error propagation, where a misclassification in one label affects subsequent predictions. Alternatively, transformer-based models that utilize contextualized semantic understanding from pre-training on large corpora have been applied to improve multi-label classification. For instance, BERT has been used for document classification~\cite{devlin2018bert,lee2021classification}. These models enhance classification robustness by leveraging the pre-trained knowledge and handling complex textual data more effectively. 

However, while BERT excels at capturing the bidirectional context of text, it does not explicitly model label correlations. To address this limitation, recent methods have integrated label co-occurrence into text encoders, improving their ability to capture inter-label dependencies and enhancing multi-label classification performance. GCLR employs contrastive learning to incorporate label graph structures into semantic representations~\cite{wang2022incorporating}. In this study, we leverage LLM prior knowledge instead of label graph information to provide a more generalizable understanding of label relationships, thereby enhancing multi-label classification tasks.

\section{Method}
\noindent \textit{Problem Statement:} We tackle a multi-label classification problem to determine the most relevant goals from the list of 17 SDGs for each aid project. This classification relies on project descriptions, meta-information such as donor and recipient countries, and the needs of different sectors and policies. Given a project description $x$ in natural language, our objective is to predict the corresponding multi-label $y$.

\begin{figure}[!t]
\centering
\includegraphics[width=0.49\textwidth]{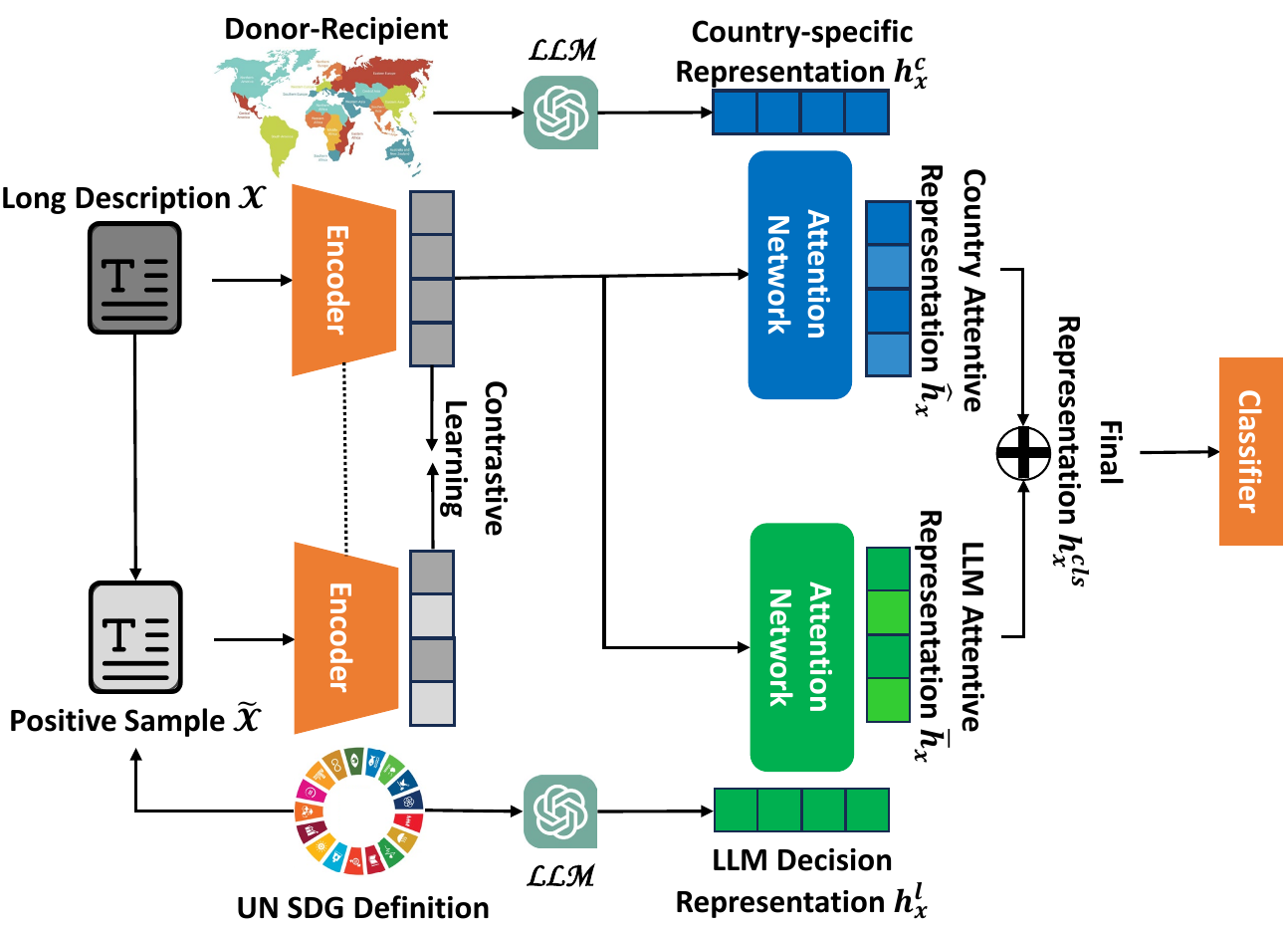} 
\caption{The proposed framework has three main components: (1) the SDG Semantics Injection Module; (2) the Country Information Guided Module; and (3) the LLM Decision Guided Module.\looseness=-1}
\label{fig:model_overview}
\end{figure}

\paragraph{Overview.} Figure~\ref{fig:model_overview} presents an overall framework, which aims to enhance the final representation $\mathbf{h}_x^{cls}$ derived from $x$ by integrating the semantic information of the SDGs into task-specific classifiers, such as BERT. The framework comprises three core modules, detailed below: 
\begin{itemize}[leftmargin=*]

\item \textbf{Semantics Injection Module ($\S$\ref{sec:sdg_info}):} Building on expert manual labeling guided by SDG definitions, we integrate the semantic representations $\mathbf{h}_g$ of the goals into the token embedding set $\mathcal{E}$ from the BERT encoder through contrastive learning. We sample essential tokens to determine aid goals and use them to construct positive samples $\Tilde{x}$ in contrastive learning.

\item \textbf{Country Information Guided Module ($\S$\ref{sec:country_info}):} Given the key information on development aid policy for donor and recipient countries, we derive an attentive representation $\hat{\mathbf{h}}_x$ that explicitly incorporates these policies to account for political contexts shaped by donor-recipient relationships.

\item \textbf{LLM Decision Guided Module ($\S$\ref{sec:llm_info}):} When the classifier encounters sentences it has not seen during training, its generalization ability, particularly for underrepresented SDGs, may be limited. We learn an attentive representation $\bar{\mathbf{h}}_x$ derived from the LLM decision. We predict the usefulness of the LLM decision $\alpha_x$ in the final representation (i.e., $h_x^{cls} =  \hat{\mathbf{h}}_x + \alpha_x\bar{\mathbf{h}}_x$).
\end{itemize}  \looseness=-1

\subsection{Semantics Injection Module}
\label{sec:sdg_info} 

During the initial phase of aid decisions, guidelines are typically established to define target objectives that align with sectoral priorities and policy needs. Aid experts then refer to these specific definitions and descriptions to determine the relevant SDG focus. 
However, this is nontrivial as projects are often framed in broad terms that emphasize overarching goals, such as eradicating extreme poverty (SDG Goal 1).

As a result, when a text classifier is trained solely with traditional supervised loss functions (e.g., binary cross-entropy), it fails to incorporate the semantic meaning of the SDGs. 
Our objective is to develop SDG-aware text representations using contrastive learning to identify latent patterns related to each SDG's semantic definition, while also extracting discriminative features that distinguish between the goals~\cite{park2022knowledge,gao2021simcse}. We use goal-semantic representations to guide the construction of positive samples $\Tilde{x}$ for contrastive learning.

\begin{table}[h!]
\begin{tcolorbox}[
  colback=black!0!white, colframe=black!20!white, colbacktitle=black!10!white, coltitle=blue!20!black ]

\footnotesize{
\textbf{Goal 1:} End poverty in all its forms everywhere \\
Target 1.a: Ensure significant mobilization of resources from a variety of sources, ...  \\

\textbf{Goal 2:} End hunger, achieve food security and improved nutrition, and promote sustainable agriculture\\
Target 2.a: Increase investment, including through enhanced international cooperation,  ... 
}
\end{tcolorbox}
\caption{Example SDG definition and  its corresponding target. \looseness=-1}
\label{fig:sdg_example}            
\end{table}

\noindent \paragraph{Positive Sample Construction.} Given  project description $x$, let $\mathcal{E}$ denote the token embedding set from BERT as follows:
\begin{align}
    \mathcal{E} = \{\mathbf{e}_1, \mathbf{e}_2, ..., \mathbf{e}_n \}
\end{align}
\noindent where $\mathbf{e}_i$ is the i-th token embedded in the BERT encoder.  We aim to construct the positive sample for $\Tilde{x}$ by sampling salient tokens conditioned on the semantic representations of the goal definition. These semantic representations are derived from the official definitions of each SDG and their corresponding targets (see Table~\ref{fig:sdg_example} for an example). The representation of the j-th goal, $\mathbf{h}_{g_j}$, can also be extracted using BERT embedding. 
The scale-dot attention weight $A$ is utilized to measure the relevance between token embedding $\mathbf{e}_i$ and goal embedding $\mathbf{g}_j$: 
\begin{align}
    Q_{i} = W_{Q}\mathbf{e}_i,~K_{j} = W_{K}\mathbf{h}_{g_j},~A_{ij} = \frac{Q_iK_j^{T}}{\sqrt{d_h}}
\end{align}
\noindent where $W_{Q}$ and $W_{K}$ are two weight matrices used for the attention mechanism. For simplicity, subscripts of all independent attention matrices are omitted. 

We sample salient tokens conditioned on the multi-label $y$ by computing the probability distribution $P$ with Gumbel-Softmax for differentiable sampling~\cite{jang2017categorical} (\eqref{eq:Gumbel}). Here, each token can contribute to more than one goal, so the importance score $P_i$ for the i-th token, with respect to the multilabel $y$ is computed by summing the probabilities of all ground-truth labels present in $y$ as follows:
\begin{align}
    P_{ij} &= Gumbel(A_{i1}, A_{i2}, , . . . , A_{i17})_j \\ \label{eq:Gumbel}
    P_{i} &= \sum_{j\in y}P_{ij}
\end{align}

Finally, we construct a positive sample $\Tilde{x}$ by retaining only the salient tokens with thresold $\tau$ and replacing all other irrelevant tokens with 0:
\begin{align}
\Tilde{x}_i = 
\left\{
\begin{array}{ll}
     x_i & \mbox{if } P_i > \tau  \\
     0 & \mbox{else } \\
\end{array}
\right.
\end{align}

\noindent \paragraph{Semantics-Aware Contrastive Learning.} The next objective is to train a text encoder that incorporates goal semantic information into its representations, using contrastive learning with the given positive sample $\Tilde{x}$. Let $h_x$ denote the average pooling of token representation of $x$ fed by BERT (i.e., $\mathbf{h}_x = \text{AvgPool}(\mathcal{E})$). Even after removing irrelevant tokens, the positive sample constructed from goal representations retains a semantic context similar to the original sentence. A shallow MLP layer, $S_{G}$, is added on top of the representation space to ensure that the sentences embedded in similar sentences are brought closer together, while the negative sample embedding $h_{x^n}$ from different sentences in the batch $\mathcal{X}$ are pushed apart through contrastive learning: \looseness=-1
\begin{align}
   \mathcal{L}_{G} &= - \log \frac{\exp(\text{sim}(S_{G}(\mathbf{h}_{x}), S_{G}(\mathbf{h}_{\Tilde{x}})))}{\sum_{x^{n}\in{\mathcal{X}}} \exp(\text{sim}(S_{G}(\mathbf{h}_{x}), S_{G}(\mathbf{h}_{x^{n}})))}
   \label{eq:s_con}
\end{align}
\noindent where \text{sim} represents the cosine similarity.

\subsection{Country Information Guided Module}
\label{sec:country_info}
Understanding the political contexts between donor and recipient countries is essential, as they expose the underlying power structures and biases that shape interdependence among multi-labels.
For example, donor's aid allocation often reflects commercial or geopolitical objectives. Similarly, each country’s SDG strategies are influenced by various institutional challenges and require a contextualized approach ~\cite{park2024addressing}.  
Thus, relying solely on semantic information from project descriptions during training is insufficient. To account for the complex doner-recipient relationships, we first extract insights derived from LLM on the countries and integrate this information using an attention mechanism to guide the text classifier.

\noindent \paragraph{Extracting  Insights on Countries.}  We design an instruction prompt to generate responses that explore the key aid policies of each donor and the development strategies of the recipient country, as in Table~\ref{fig:country_prompt}. We include the income level of the recipient country in prompts because of a significant connection between the two variables~\cite{knack2011aid,EasterlyWilliam2014}. The income group, which best represents aid demand and absorptive capacity of a recipient country, serves as a primary criterion for effective allocation of aid~\cite{COLLIER20021475,Dalgaard2004,Easterlyetal2008}. For example, for lower-income countries, the focus of aid tends to be on poverty alleviation, basic needs, and infrastructure. Middle-income countries may prioritize advanced developmental outcomes such as institutional development, technological innovation, and economic governance for growth.

\begin{table}[t!]
\begin{tcolorbox}[
  colback=black!0!white, colframe=black!20!white, colbacktitle=black!10!white, coltitle=blue!20!black ]

\footnotesize{
Based on government documents, summarize the Official Development Assistance (ODA) Policy of [donor country]?}
\end{tcolorbox}

\begin{tcolorbox}[
  colback=black!0!white, colframe=black!20!white, colbacktitle=black!10!white, coltitle=blue!20!black ]

\footnotesize{
In the OECD CRS (Creditor Reporting System) database, income groups categorize countries based on their gross national income (GNI) per capita. Here’s what each group represents: \\ 
- LICs (Low-Income Countries): Countries with a GNI per capita of \$1,045 or less. \\
- LMICs (Lower Middle-Income Countries): Countries with a GNI per capita between \$1,046 and \$4,095. \\
- UMICs (Upper Middle-Income Countries): Countries with a GNI per capita between \$4,096 and \$12,695. \\
- LDCs (Least Developed Countries): This group includes countries identified by the United Nations as facing severe structural impediments to sustainable development. 

Based on credible government papers, can you summarize the Official Development Assistance (ODA) Policy of [income classification], [recipient country]?}
\end{tcolorbox}
\caption{Prompt example used in research. The “[country]"
part will be replaced by the country name, and  “[income classification]" will indicate the recipient country's GNI per capita level. 
\looseness=-1}
\label{fig:country_prompt} 
\end{table}

\noindent \paragraph{Country Information-Driven Attention.} Given text input $x$ and its corresponding model responses regarding donor and recipient countries, we extract the donor representation $\mathbf{h}_x^{d}$ and the recipient representation  $\mathbf{h}_x^{r}$  from BERT encoder. The country-specific representation $\mathbf{h}_x^{c}$ is then constructed by combining the donor and recipient representations through element-wise multiplication (see \eqref{eq:h_xc}). To incorporate this country information into the text representation, we introduce cross-attention between the token embedding set $\mathcal{E}$ and the country representation $\mathbf{h}_x^{c}$ as follows: 
\begin{align}
    \mathbf{h}_x^c &= S_C(\mathbf{h}_x^{d} \odot \mathbf{h}_x^{r}) \label{eq:h_xc}\\ 
    Q_{i} &= W_{Q}\mathbf{e}_i,~K = W_K\mathbf{h}_x^{c},~A_{i} = \frac{Q_iK^{T}}{\sqrt{d_h}} \\
    \hat{A} &= \text{Softmax}(A) \\
    \hat{\mathbf{h}}_x &=  \text{AvgPool}(\{\hat{A}_1\mathbf{e}_1, \hat{A}_2\mathbf{e}_2, ..., \hat{A}_n\mathbf{e}_n\}) \label{eq:country_h} 
\end{align}
\noindent where $S_C$ is the MLP layer and AvgPool is the average pooling applied to the token representations.

We append an auxiliary classifier $f_c$ to the attentive representation $\hat{\mathbf{h}}_x$ with the conventional binary cross-entropy loss $H$ to directly perform multi-label classification.  The refined representation $\hat{\mathbf{h}}_x$ 
plays an important role in deciding on
which tokens to focus on based on the country information.
\begin{align}
    \mathcal{L}_C = H(y, f_c(\hat{\mathbf{h}}_x))
    \label{eq:country_loss}
\end{align}

\subsection{LLM Decision Guided Module \looseness=-1}
\label{sec:llm_info}
The CRS is the official databse that incorporates input from a wide range of countries and experts on international aid. However, this inclusivity leads to varied syntactic and semantic distributions in the textual data, presenting significant challenges for multilabel classifiers. Limited training data for underrepresented SDGs further limit the generalizability of task-specific classifiers. In contrast, LLMs are known to demonstrate broader generalization capabilities due to their higher representational capacity, which we integrate using an attention mechanism.

\noindent \paragraph{Decision-Driven Attention.} We first extract the LLM decision $\bar{y}$ for the project description $x$ by building a prompt using the definition of each goal, as illustrated in Table~\ref{fig:llm_decision}. The decision representation $\mathbf{h}_x^{l}$ is constructed by sampling the goal representation corresponding to the decision $\bar{y}$ combining them through multiplication, and passing the result through an MLP layer $S_L$  (\eqref{eq:h_xl}). Similarly to incorporating country information, we applied a mechanism of cross-attention between the token embedding set and the decision representation as follows: \looseness=-1
\begin{align}
\mathbf{h}_x^{l} &= S_L(\prod\limits_{j \in \bar{y}} \mathbf{h}_{g_j}) \label{eq:h_xl}\\
 Q_{i} &= W_{Q}\mathbf{e}_i,~K = W_K\mathbf{h}_x^{l},~A_{i} = \frac{Q_iK^{T}}{\sqrt{d_h}} \\
 \bar{A} &= \text{Softmax}(A) \\
    \bar{\mathbf{h}}_x &=  \text{AvgPool}(\{\bar{A}_1\mathbf{e}_1, \bar{A}_2\mathbf{e}_2, ..., \bar{A}_n\mathbf{e}_n,\})
\end{align}

Consistent with \eqref{eq:country_loss}, we apply  standard binary cross-entropy loss using the auxiliary classifier $f_l$ on top of the attentive
representation $\bar{h}_x$:
\begin{align}
    \mathcal{L}_D = H(y, f_l(\bar{\mathbf{h}}_x))
    \label{eq:decision_loss}
\end{align}

\begin{table}[t!]
\begin{tcolorbox}[
  colback=black!0!white, colframe=black!20!white, colbacktitle=black!10!white, coltitle=blue!20!black ]
\footnotesize{
Now you will help me classify the sentence according to the following 17 Sustainable Development Goals (SDGs): \\
\textbf{Goal 1:} End poverty in all its forms everywhere \\
... \\
$<\text{EXAMPLES}>$:  [Project description]

Determine which of the above Sustainable Development Goals (SDGs) does the $<\text{EXAMPLES}>$ correspond to. You can select multiple answers following format. \\ 
$<\text{Answer}>$: $<1 \sim 17>$
}
\end{tcolorbox}
\caption{Prompt to extract the language model decision for each project description: The “[Project description]"
part will be filled with the specific project description. \looseness=-1}
\label{fig:llm_decision}         
\end{table}

\noindent \paragraph{Decision Evaluation.} Naive utilization of  LLM decision can result in suboptimal prediction performance. To address this issue, we propose an additional evaluation process that assesses the usefulness of decision and calibrates their weights for subsequent multi-label predictions. We generate the usefulness label $y_{u}$ by determining whether the original label $y$ overlaps with the decision $\bar{y}$. We evaluated the effectiveness of the decision-driven representation of the model $\bar{\mathbf{h}}_x$ using this usefulness label $y_{u}$. We train the usefulness classifier $f_u$ with binary cross-entropy loss:
\begin{align}
    \mathcal{L}_U = H(y_u, f_u(\bar{\mathbf{h}}_x))
    \label{eq:useful_loss}
\end{align}

\noindent \paragraph{Multi-label Prediction with Calibrated Representation.} 
Using the usefulness classifier $f_u$ , we determine the scaling coefficient $\alpha_x$ by evaluating the usefulness of the LLM decision-driven representation $\bar{\mathbf{h}}_x$ based on the prediction probability of the classifier (that is, $\alpha_x=f_u(\bar{\mathbf{h}}_x)$). The refined final representation $h_x^{cls}$ is derived by combining the country information-driven representation $\hat{h}_x$ with the calibrated decision-driven representation $\bar{\mathbf{h}}_x$   (\eqref{eq:h_cls}). Then the final representation $\mathbf{h}_x^{cls}$ is then passed through the MLP classifier $f_{cls}$ for final classification prediction:
\begin{align}
    h_x^{cls} &=  \hat{\mathbf{h}}_x + \alpha_x\bar{\mathbf{h}}_x \label{eq:h_cls} \\
    \mathcal{L}_{CE} &= H(y, f_{cls}(\mathbf{h}_x^{cls}))
    \label{eq:bce_loss}
\end{align}

Finally, the overall loss function is computed by combining the loss terms discussed above with hyperparameters $\lambda_1, \lambda_2$  as given in \ref{eq:total_loss}.
\begin{align}
    \mathcal{L}_{total} = \mathcal{L}_{CE} + \lambda_{1}\mathcal{L}_G + \lambda_{2}(\mathcal{L}_C + \mathcal{L}_D + \mathcal{L}_U)
    \label{eq:total_loss}
\end{align}
For inference, we extract country information and decision for the test data $x_t$. We then perform an element-wise summation of the refined characteristics: the country information-driven representation $\hat{\mathbf{h}}_{x_t}$ and the calibrated decision-driven representation $\alpha_{x_t}\bar{\mathbf{h}}_{x_t}$ to obtain the final output characteristic (i.e., $h_{x_t}^{cls} =  \hat{\mathbf{h}}_{x_t} + \alpha_{x_t}\bar{\mathbf{h}}_{x_t}$). The
final feature is then passed to the classifier $f_{cls}$.
\looseness=-1

\section{Experiment}
\subsection{Performance Evaluation}
\noindent \textbf{Dataset.} 
As introduced in Section 1, our main dataset, the CRS is internationally recognized as the standard for detailed project-level aid data, providing breakdowns by country, sector, type of aid, donor, and year. Based on consultations with aid experts, among the 93 data fields, we selected four key features: \textit{project description}, \textit{SDG focus index}, \textit{donor code}, and \textit{recipient code}. The \textit{project description} provides a detailed text of up to 4,000 words, used to classify the project sector and its alignment with the aid goal. Introduced in 2018,  \textit{SDG focus index} is a multi-entry label indicating up to ten SDGs the project focuses on. However, reporting remains voluntary due to the significant manpower required for data entry and review~\cite{dcd2018a}.

\begin{figure}[!t]
\centering
\includegraphics[width=0.47\textwidth]{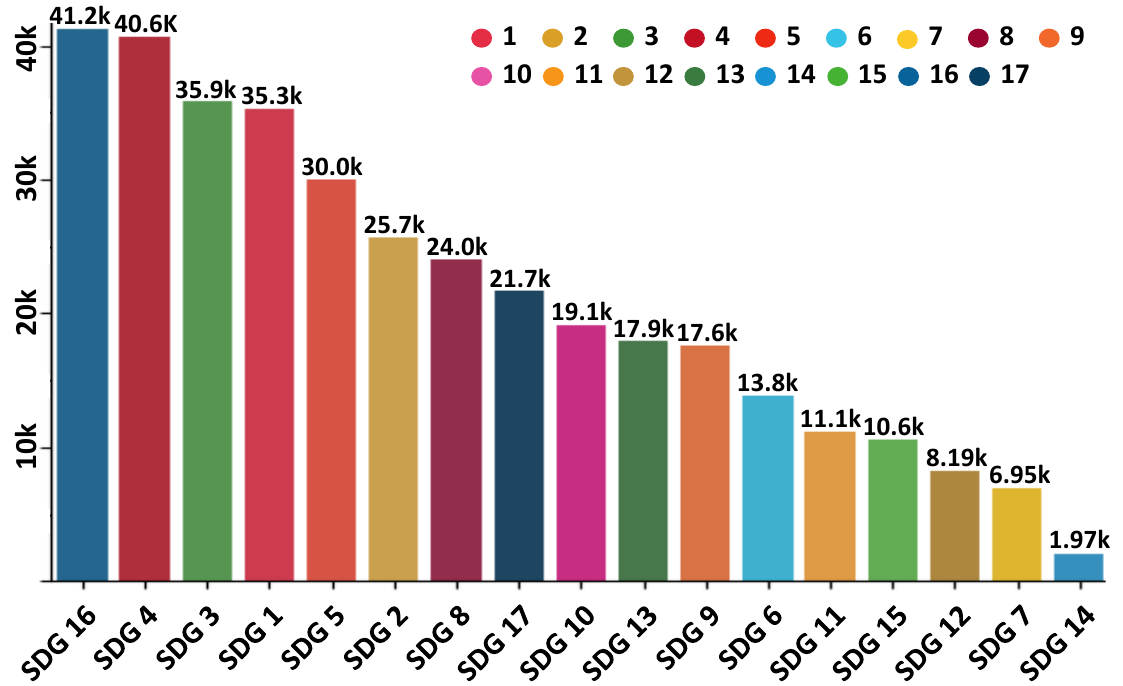} 
\caption{Distribution of SDG focus index by frequency ($y$).\looseness=-1}
\label{fig:sdg_statics}
\end{figure}

The labels $y$ correspond to the 17 SDGs listed in this focus index. Figure~\ref{fig:sdg_statics} shows the data statistics, indicating underrepresented goals such as SDG 7 and SDG 14. The \textit{donor code} and \textit{recipient code} are the nations participating in each aid project. The former is a numerical value, often one of the 32 member countries of the OECD Development Assistance Committee (DAC) or a multilateral donor organization. The latter assigns a numerical value to the 170 recipient countries or territories that receive international aid.

\noindent \textbf{Implementation.} We gathered 151,832 official aid documents and divided them into training and testing sets in a 3:1 ratio. This division was independent of the submission year, ensuring consistency across the data's studied fields. Due to the bilingual nature of the documents (some in English, others in French), we utilized a multilingual BERT as the foundation and trained LLMs over 100 epochs. The Adam optimizer for BERT was configured with a learning rate of 1e-05. The hyperparameters  in~\ref{eq:total_loss} were set to 0.1, with a token sampling threshold $\tau$ of 0.01. Text-based baselines were evaluated using the same backbone network, optimizer, and learning rate. For graph-based baselines, the adjacency matrix of the label graph was derived from the co-occurrence matrix, which was constructed by counting the frequency of label pairs in the training dataset. This label information was then incorporated into the text encoder.
We used OpenAI's GPT-4o-mini with default settings of the top-p value at one and the temperature at zero. 

\noindent \paragraph{Result.} We evaluated our framework against two LLM-based classifiers and three text encoder-based classifiers.
\begin{itemize}[leftmargin=*]
\item  \textsf{GPT-4o} utilizes in-context learning for inference.
\item \textsf{GPT-4 Turbo} uses in-context learning for inference.
\item \textsf{Vanilla BCE} is a text encoder-based baseline that applies conventional binary cross-entropy for multi-label classification.
\item \textsf{LGCN} integrates label graph information into the text encoder using an attention~\cite{ma2021label}. 
\item \textsf{GCLR} utilizes contrastive learning to integrate label graph information into the text encoder~\cite{wang2022incorporating}.
\end{itemize}

Table~\ref{tab:result} reports the performance of multi-label classifiers, showing that our framework consistently surpasses baseline models across all metrics. The table also confirms that text encoder-based classifiers generally perform better than LLM-based classifiers, suggesting the importance of having contextual information.

\begin{table}[!t]
\centering
\scalebox{0.9}{
\begin{tabular}{l|cccc}
\toprule
Model & Precision & Recall & F1-score & AUROC \\ \midrule
GPT-4o & 0.7409	& 0.7578	& 0.7170 &	- \\
GPT-4 Turbo & 0.7407 & 0.8049 & 0.7381 & - \\
Vallia BCE & 0.8509 & 0.8004 & 0.8243 & 0.9382 \\
LGCN & 0.8628 & 0.8075 & 0.8365 & 0.9521 \\
GCLR & 0.8572 &	0.8142 &	0.8357 & 0.9519\\ \midrule
Ours & \textbf{0.8684} & \textbf{0.8192} & \textbf{0.8430} & \textbf{0.9617} \\ \bottomrule 
\end{tabular} 
}
\caption{Performance comparison of multi-label classification methods on the CRS dataset. Best result is shown in bold text.}
\label{tab:result}
\end{table}

\noindent \paragraph{Ablation Study.} 
We conducted ablation experiments in which each component was removed from the entire model and considered four configurations:
\begin{itemize}[leftmargin=*]
\item  \textsf{Full Components}: Our complete method.
\item  \textsf{w/o SDG Semantic}: The method excluding contrastive learning loss $L_{G}$, as detailed in \eqref{eq:s_con}.
\item  \textsf{w/o Country Info }: The approach without employing the country information-based attention in \eqref{eq:h_xc} - \ref{eq:country_loss}.
\item \textsf{w/o LLM Decision}: The method omitting the language model-specific attention, referenced in \eqref{eq:h_xl} - \eqref{eq:h_cls}.
\end{itemize}

\begin{table}[!t]
\centering

\scalebox{0.9}{
\begin{tabular}{l|cccc}
\toprule
Model & Precision & Recall & F1-score & AUROC \\ \midrule
Full Component & 0.8684 & \textbf{0.8192} & \textbf{0.8430} & \textbf{0.9617} \\
w/o SDG Semantic & 0.8631 & 0.8121 & 0.8368 & 0.9588 \\
w/o Country Info & 0.8671 & 0.8143 & 0.8399 & 0.9594 \\
w/o LLM Decision & \textbf{0.8787} & 0.7971 & 0.8359 & 0.9609 \\ \bottomrule
\end{tabular} 
}
\caption{Ablation study results. The removal of any model component results in a reduction of F1 score and AUROC.}
\vspace*{-3mm}
\label{tab:ablation}
\end{table}

\begin{figure*}[t!]
\centerline{
      \includegraphics[width=0.98\linewidth]{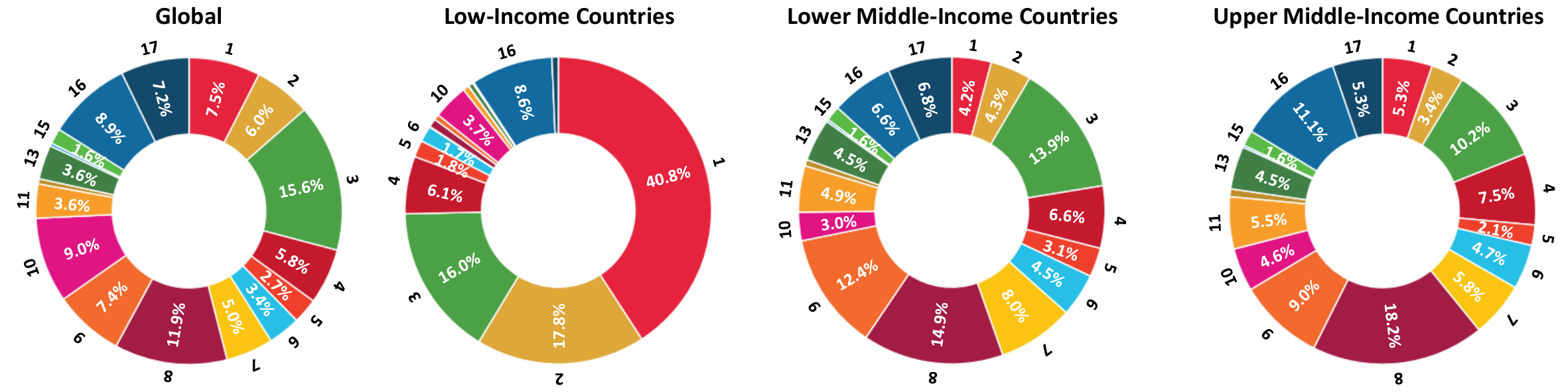}}
      \caption{Distribution of Estimated SDG Budget Allocation by GNI per Capita. SDGs with a ratio of less than 1\% have been omitted. \looseness=-1
      } 
\label{fig:sdg_pie}
\end{figure*}

\begin{figure}[!t]
\centering
\includegraphics[width=0.46\textwidth]{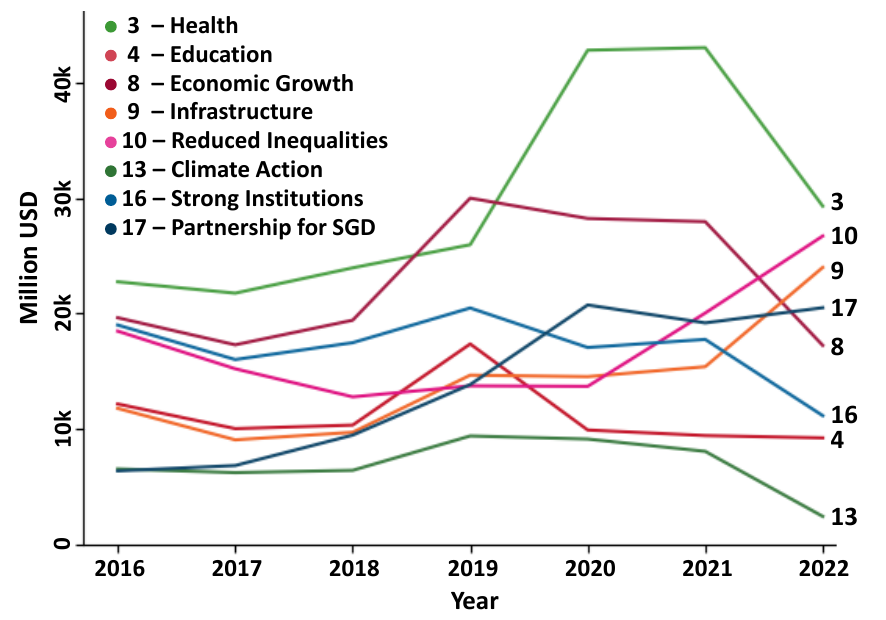} 
\caption{Estimated annual budget allocation. Over a 7-year period for a subset of SDGs with the most variability.\looseness=-1}
\label{fig:sdg_year}
\end{figure}

Table~\ref{tab:ablation} shows the performance for each ablation. Each component positively contributes to the improvement of classification performance. The model without the LLM decision achieves higher precision but lower recall, indicating that the SDG Semantics Injection Module and the Country Information Guided Module play a crucial role in enhancing precision, while incorporating the language model's decision further boosts recall. In particular, the LLM decision improved recall for underrepresented classes such as Goal 7 (0.74 $\rightarrow$ 0.81) and Goal 14 (0.68 $\rightarrow$ 0.73). Due to space limitations, we show results on additional analyses such as the LLM knowledge qualitative analysis and sensitivity analysis for alternative LLM backbones in the Appendix.

\section{Implication}
We proposed a practical method to map aid projects with the SDGs. By capturing a comprehensive range of project dimensions, our model could improve classification accuracy. Our method is recognized as a substantial contribution to the fields of international development, public policy, and financial management, allowing a more precise evaluation of global development efforts. Moreover, it supports data-driven decision making (DDDM), allowing policy makers to systematically prioritize focus areas and formulate evidence-based policies. We discuss implications for tracking and monitoring aid projects and how AI methods can be used in the strategic planning and execution of development projects. \looseness=-1
\subsection{Discussion} 
For financial analysis, we imputed all data from 2015 to 2022 for missing historical aid projects using our model.

\noindent \paragraph{Limitations in Analyzing Past Data.} 
As aid categorization has been voluntarily recorded since 2018, there exist no labels prior to that year, and a significant portion of the SDG category data remain missing even after 2018. Much of the data after 2018 suffer from incomplete labels, leading to analyses that may not accurately reflect the actual aid allocations and impacts (see Figure 6 in the Appendix). This issue limits the historical analysis of contributions.

\noindent \paragraph{Analysis of Aid Financing Using the Trained Model.} To estimate labels for each past and missing project, we first extracted donor-recipient information and decisions through an language model, then applied this information, along with project descriptions, to our trained model. Although a single project may align with multiple goals, distributing the budget equally across all SDGs is suboptimal. We assume that the annual budget sum $S_i$ for the $i$-th year in the labeled dataset can be expressed as a linear sum of the statistics of the $i$-th year for each $k$-th goal, $c_i^k$, within the labeled dataset. To estimate the budget $w_k$ for each $k$-th goal,  we fit a linear regression model by solving the optimization problem (\eqref{eq:optimization}). The estimated budget proportions, detailed in the Appendix Table 6, are then used in the financing analysis.
\begin{align}
    \min_{w_1, ... , w_{17}} \sum_{i \in \{2018,..,2022\}} || S_i - \sum_{k = 1}^{17} (c_k^i \times w_k)  ||^2_2   \label{eq:optimization} 
\end{align}

Our analysis highlights distinct trends from both the recipient and donor perspectives. 
From the recipient side, illustrated in Figure~\ref{fig:sdg_pie}, the allocation of the aid varies by income group. Low-income countries receive concentrated support for foundational goals like SDG 1 (No Poverty) and SDG 2 (Zero Hunger), while lower and upper middle-income countries see increased allocations for goals like SDG 8 (Decent Work and Economic Growth) and SDG 9 (Industry, Innovation, and Infrastructure). This indicates that development needs and economic status shape the distribution of aid, emphasizing the importance of aligning aid with the priorities of different income groups.

Figure~\ref{fig:sdg_year} shows how donor priorities have shifted across goals, particularly in response to global events over the course of seven years. For example, funding for SDG 3 (Good Health and Well-being) has nearly doubled since 2018 due to the COVID-19 pandemic, before returning to pre-pandemic levels by 2022. This fluctuation demonstrates donors' capacity to adjust priorities in response to urgent global needs.
Methods like ours help better understand international development financing, offering insights that were previously unattainable due to extensive missing data and the lack of structured SDG-specific data in aid records.

\subsection{Expert Interview}

We conducted three semi-structured interviews and two focus group discussions to identify stakeholder's challenges and assess our AI model's relevance and effectiveness. A total of eight aid experts participated in the study, belonging to different government agencies. Further details about the experts are provided in the Appendix.

The interviews revealed three main challenges in the allocation of SDGs: categorization difficulties in multifaceted projects, variations in expertise and priorities between the personnel involved in the assignment, and the influence of policy priorities. Addressing these challenges requires nuanced project classification and standardized evaluation criteria.

Efficacy of the AI model could be summarized into three parts: efficiency, standardization, and potential. 
First, experts found that the use of AI model in assigning the aid goal to each project reduces time and increases the precision of their given task. 
For example, the multidimensional nature of development projects, especially those that target various objectives such as energy efficiency and climate change mitigation, has been adding complexities in assigning multiple SDGs.  
Experts identified that model recommendations can act as a guideline that reduces human labor and increases accuracy by giving more time to review the answers rather than solving them, indicated in the quote below: 
\begin{quote}
    \textit{“It could significantly reduce administrative workload, increasing efficiency, especially for those reviewing numerous projects annually. (OPC)”}
\end{quote}

Second, experts found that our AI model can enhance credibility and coherence by minimizing human subjectivity and political influence, offering a standardized framework for aid classification and policy evaluation. This capability, in particular, addresses stakeholders' concerns about inconsistencies, arbitrariness in officials' interpretations, and the lack of clear guidelines, all of which threaten data quality, as quoted:
\begin{quote}
    \textit{“Automatic SDG assignment provides a standardized guideline that reduces subjective variations and greatly helps establish uniformity. (KOICA)”}
\end{quote}

Third, the model has the potential to support the efforts of the OECD and the UN to align aid statistics with the SDGs by encouraging broader participation in completing the currently optional \textit{SDG focus index} field. By simplifying this process, it can promote wider participation among governments and strengthen their commitment to sustainable development.

\subsection{Conclusion}
We presented a model that uses LLM to classify international aid projects into SDG categories. This provides practical implications, assisting project managers in the time-consuming task of manually assigning multiple SDGs and supporting government experts in reviewing annual aid data. Our approach could potentially lead to a revision of the CRS guidelines, making SDG categorization mandatory and improving data quality and completeness, thus improving the reliability of development assessments.

We introduced three components to address the challenges in the classification of aid projects. Analytically, we compiled a new dataset that enriches existing aid information. The model shows potential for conducting detailed retroactive analyses of aid financing trends and donors' behavior, enabling a critical assessment of whether aid activities align with SDG commitments and meet recipients' demand.

The integration of AI in aligning aid projects with goal priorities offers significant theoretical implications. Our study serves as a salient example of how innovative technology can contribute to global governance for sustainable development, offering a foundation for further research. The expert interview also suggests that, by improving scalability, transparency, and aid effectiveness, our approach advances the SDG reporting system and informs future efforts to improve data-driven decision-making in international development. 

\section*{Acknowledgments} S. Park and D. Lee contributed equally as co–first authors. M. Cha and K.R. Park are co–corresponding authors.  This
work was supported by the Center for Science Technology
for Global Development (CSTG) at Korea Advanced Institute of Science and Technology (KAIST) Research Program
(Grant Number: A0601006006) and the National Research
Foundation of Korea (2022S1A5A8052217).
\bibliographystyle{named}
\bibliography{ijcai25}

\newpage

\appendix
\section{APPENDIX}

\subsection{LLM Knowledge Qualitative Analysis} 

Using language models in multi-label classification yields both encouraging results as well as limitations. Its prior knowledge demonstrated significant accuracy in its representations of recipient countries when associated with income group categorizations. For example, the model responses in Low-Income Countries frequently emphasized basic developmental priorities such as necessary services and poverty alleviation, while those in Middle-Income Countries placed a proper emphasis on institutional development and economic governance in Figure~\ref{fig:llm_reponse_example}. These categorizations, which are fundamental to the CRS database and international development policy frameworks, revealed the language model's ability to construct coherent, contextually appropriate profiles based on recipient income classifications.

For donor nations, the model showed a reasonable understanding of basic policy orientations and common aid priorities, accurately reflecting shared patterns even as national ODA policies changed. However, the study identified particular discrepancies between LLM's responses and actual policy details. For example, in Japan, the language model emphasized the inclusion of gender and women's empowerment, both of which are key themes in the new 2023 strategy. However, the model did not reflect the aims of Japan's previous ODA framework, which was more focused on quality growth and human security, indicating a significant gap in temporal alignment with real policy. Similarly, the model overlooked a key feature of Korea's approach to development aid: each receiving country has an independently crafted Country Partnership Strategy. These findings imply that, while the language model could serve as a useful baseline for country-specific ODA situations, it is critical to incorporate the knowledge of political experts.

\begin{table}[!t]
\setlength{\tabcolsep}{3.5pt}
\centering

\scalebox{1.0}{
 \begin{tabular}{c|c|c} \toprule
 \textbf{SDG} &  \textbf{Description} & \textbf{Prop.} \\ 
 \toprule
 8 & Decent Work and Economic Growth & 1.00  \\
 3 &Good Health and Well-being & 0.97\\
 16 & Peace, Justice, and Strong Institutions & 0.60  \\ 
2 & No Poverty & 0.41 \\
 15 & Life on Land &  0.39 \\ \hline
9 & Industry, Innovation, and Infrastructure & 0.36 \\
 12 & Responsible Consumption and Production & 0.34 \\ 
 10 & Reduced Inequalities & 0.33 \\
  11 & Sustainable Cities and Communities & 0.29 \\
 17 & Partnerships for the Goals & 0.26 \\ \hline
 7 & Affordable and Clean Energy & 0.25  \\ 
 14 & Life Below Water & 0.24 \\
  13 & Climate Action & 0.20 \\
 5 & Gender Equality & 0.18 \\
 6 & Clean Water and Sanitation & 0.11  \\  \hline
 1 & Zero Hunger & 0.06  \\ 
 4 & Quality Education & 0.02  \\ 

 \bottomrule
\end{tabular} 
}

 \caption{
SDGs and Estimated Budget Proportions from \eqref{eq:optimization}}
 \label{tab:SDG_definition}
\end{table}

\begin{figure}[t!]
\begin{tcolorbox}[
  colback=black!0!white, colframe=black!20!white, colbacktitle=black!10!white, coltitle=blue!20!black ]
  LICs: The Development Policy of the Syrian Arab Republic aims to reduce poverty through various initiatives. This includes investing in infrastructure, education, healthcare, and social welfare programs to improve the quality of life for its citizens. \\
  LMICs: Nigeria's development policy as an LMIC has outlined key priorities in its development policy, including investing in infrastructure development, promoting industrialization and diversification of the economy.  \\
 UMICs:  Chile's development policy for UMICs focuses on promoting sustainable development, and improving social inclusion through targeted policies and programs.
\end{tcolorbox}
\captionof{figure}{Example Summarization of LLM Development Policy Knowledge Based on the Recipient Country’s Development Status. \looseness=-1}
\label{fig:llm_reponse_example} 
\end{figure}
\begin{figure*}[t!]
\centerline{
      \includegraphics[width=1\linewidth]{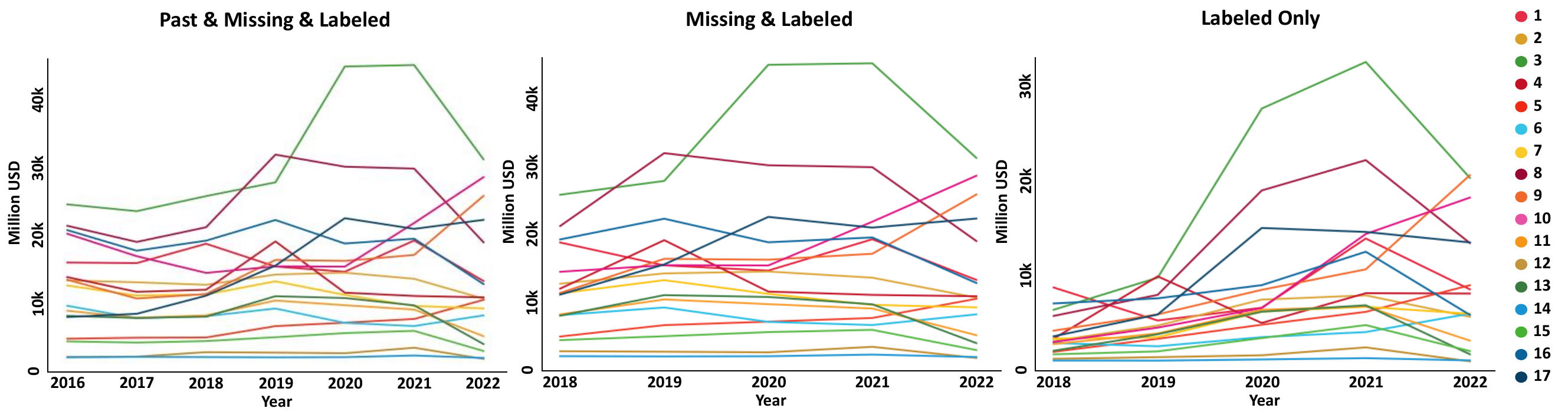}}
      \caption{Estimated Annual Budget Allocation for Each SDG. The full dataset, which includes past and missing data, exhibits a different distribution compared to the labeled dataset alone. 
      } 
\label{fig:sdg_year_entire}
\end{figure*}
\vspace{1cm}
\begin{figure*}[t!]
\centerline{
      \includegraphics[width=1\linewidth]{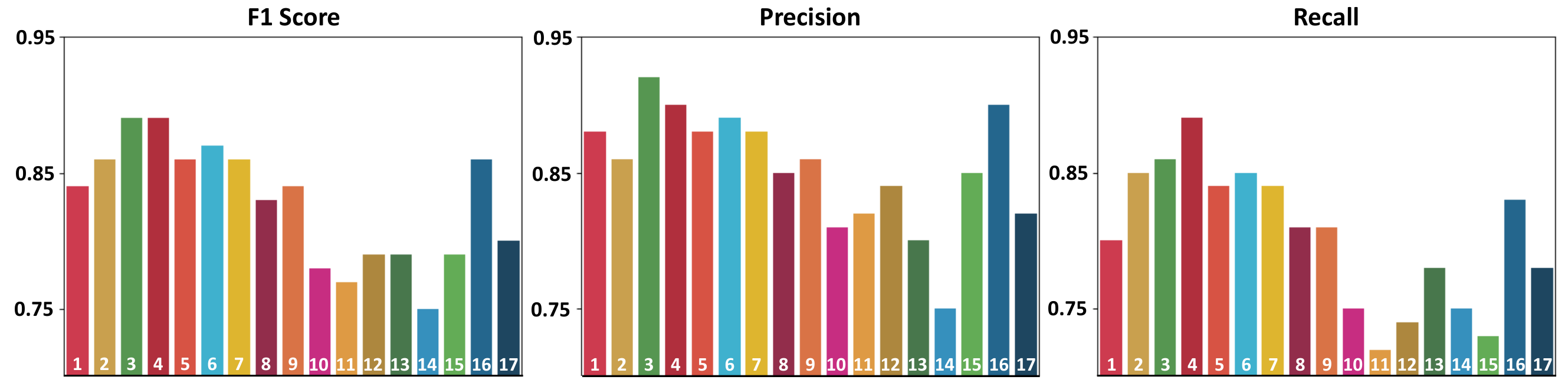}}
      \caption{Detailed performance metrics, including F1-score, precision, and recall, for each SDG. \looseness=-1
      } 
\label{fig:peromance_metric}
\end{figure*}

\begin{table}[!t]
\centering

\scalebox{1.0}{
\begin{tabular}{l|cccc}
\toprule
Model & Precision & Recall & F1-score & AUROC \\ \midrule
Gemini-Pro & 0.8672 & 0.8177 & 0.8417 & 0.9617 \\
GPT-3.5 & 0.8672 & 0.8188 & 0.8423 & 0.9617 \\
GPT-4o & 0.8668 & 0.8159 & 0.8406 & 0.9616 \\
GPT-4-turbo & 0.8684 & 0.8167 & 0.8417 & 0.9617 \\\bottomrule
\end{tabular} 
}

\caption{Performance comparison by different LLM backbones. }
\label{tab:llm_decision}
\end{table}

\subsection{Implementation Details and Results}

\subsubsection{Dataset} 

All datasets used in this paper are non-proprietary online resources that are easily accessible: \url{data-explorer.oecd.org/}. The criteria for creating the CRS dataset were established by the OECD DAC Working Party on Development Finance Statistics. For the SDG goal and target definitions, we use the formal definitions provided by the UN. Originally, the \textit{SDG focus index} consists of 17 goals and 169 targets, this study focuses exclusively on the goal level, as the targets are subordinate components of the goals.

The training data set was constructed using data from 12 donor countries, selected based on two criteria: (1) the degree to which they completed the SDG focus index in their reporting, and (2) their membership in the International Aid Transparency Initiative (IATI), an organization dedicated to promoting transparency in development resource allocation and outcomes.  The donor countries that were included in the training datasets are Canada, Denmark, EU, Finland, France, Germany, Iceland, Ireland, Italy, Japan, Korea, Luxembourg, Netherlands, and Norway. During the inference process, data encompassing all donor and recipient codes was used. We analyze a total of 1,719,733 datasets spanning the years 2015 to 2022, the period since the definition of the SDGs. This includes 382,083 historical records that preceded the establishment of voluntary input criteria in 2018, as well as 616,726 entries with missing data post-2018.

\subsubsection{Implementation} 

To ensure consistency in results, experiments replicating existing baselines were performed with fixed settings for learning, backbone networks, and other relevant parameters. 
For baseline models utilizing label graph information, we first constructed a concurrence matrix by counting the co-occurrences of label pairs in the training dataset. Next, we calculate the frequency of occurrence of each label to create a conditional probability matrix. Then a threshold of 0.1 was applied to the conditional probabilities to construct the graph. 

\subsubsection{Result} 

Figure~\ref{fig:sdg_year_entire} illustrates the estimated annual budget allocation for each SDG, categorized by the volume of data analyzed. Relying solely on a labeled dataset risks producing results biased toward specific countries, limiting the comprehensiveness of insights that could be derived from analyzing the full dataset.
Figure~\ref{fig:peromance_metric} presents the F1-score, precision, and recall for each SDG.  SDGs 1, 3, 4, and 16, which are supported by extensive datasets, exhibit higher performance metrics. In contrast, SDGs 7, 12, 14, and 15, constrained by limited datasets, show notably lower performance, particularly in recall. This disparity underscores the importance of data development and augmentation efforts to improve the representation and performance of underrepresented SDGs.

\newpage
\subsubsection{Sensitive  Analysis}  

During the inference stage, we extract decisions from the LLM for the evaluation data. Furthermore, we performed a sensitivity analysis testing our method on four alternative LLM backbones whose decisions were not utilized during training. The results in Table~\ref{tab:llm_decision} show only a marginal performance loss, indicating that the model remains robust, even when the LLM decision used during inference differs from the one used during training.

\subsection{Expert Interview Details} 

We conducted three semi-structured interviews and two focus group discussions to enhance our understanding of the CRS reporting process, identify stakeholder challenges, and assess our AI model's relevance and effectiveness. Eight key stakeholders from four government agencies involved in ODA statistics: the Office for Government Policy Coordination (OPC), the Korea International Cooperation Agency (KOICA), the Export-Import Bank of Korea (KEXIM), and Statistics Korea, were selected for their relevance to ODA and SDGs classification in their daily job responsibilities.

\end{document}